\documentstyle[epsf,iopconf1]{article}

\newcommand{\be}[1]{\begin{equation} \label{(#1)}}
\newcommand{\ee}{\end{equation}}
\newcommand{\ba}[1]{\begin{eqnarray} \label{(#1)}}
\newcommand{\ea}{\end{eqnarray}}

\begin{document}

\title{The Quest for the Neutrino Mass Spectrum}

\author{H.V. Klapdor--Kleingrothaus, H. P\"as}

\section{Introduction}
Recently the particle physics community was shocked with breathtaking 
news from the neutrino sector: Neutrino oscillations have been 
confirmed finally in the Super-Kamiokande \cite{superk}
experiment. 
Now for the first time, ongoing and future experiments in 
neutrino oscillations (Super-Kamiokande, Borexino, SNO, MINOS, KAMLAND,
MINIBOONE,...) and
double beta decay 
(Heidelberg--Moscow, GENIUS,...)
together can aim to solve the neutrino mass puzzle.

It was in 1930, when  Wolfgang Pauli (fig. \ref{wp})
wrote his famous letter adressed 
as {\it Liebe radioaktive Damen und Herren
(Dear radioactive Ladies and Gentlemen)}, where he informed the 
participants of a nuclear physics workshop in T\"ubingen about his absence
(he preferred to participate in a dance party) and postulated the 
neutrino to solve the problem of energy nonconservation in the nuclear beta 
decay. In 1956 the neutrino was observed for the first time by Clyde Cowan 
and 
Fred Reines in Los Alamos, 
who originally planned to explode a nuclear bomb for 
their experiment \cite{reines}.
Finally, two years ago, the Super-Kamiokande experiment,
a 50,000 ton water tank viewed by more than 11,000 photo multipliers 1,000
meter underground below a holy mountain in Japan,
announced a significant signal for neutrino oscillations and established
a non-vanishing 
mass of the neutrino as the first experimental signal of physics beyond 
the standard model.
However, in spite of these successes, entering a new millenium 
the neutrino is still the most mysterious of the known particles. 
Alternatingly compared with spaceships travelling through
the universe, ghosts 
penetrating
solid rocks and vampires missing a mirror image \cite{sutton}, 
it still inspires the 
phantasy of hundreds of adventurous particle, nuclear and astro physicists
being motivated by the hope, the neutrino could act as a key to the old 
human dream of a final theory, describing all particles and forces in a 
unified framework, and to a deeper understanding of the fate of the universe.

The attributes, making the neutrino this kind of outlaw among the known 
particles, are the following:

\begin{figure}
\parbox{4cm}{
\epsfxsize=40mm
\epsfbox{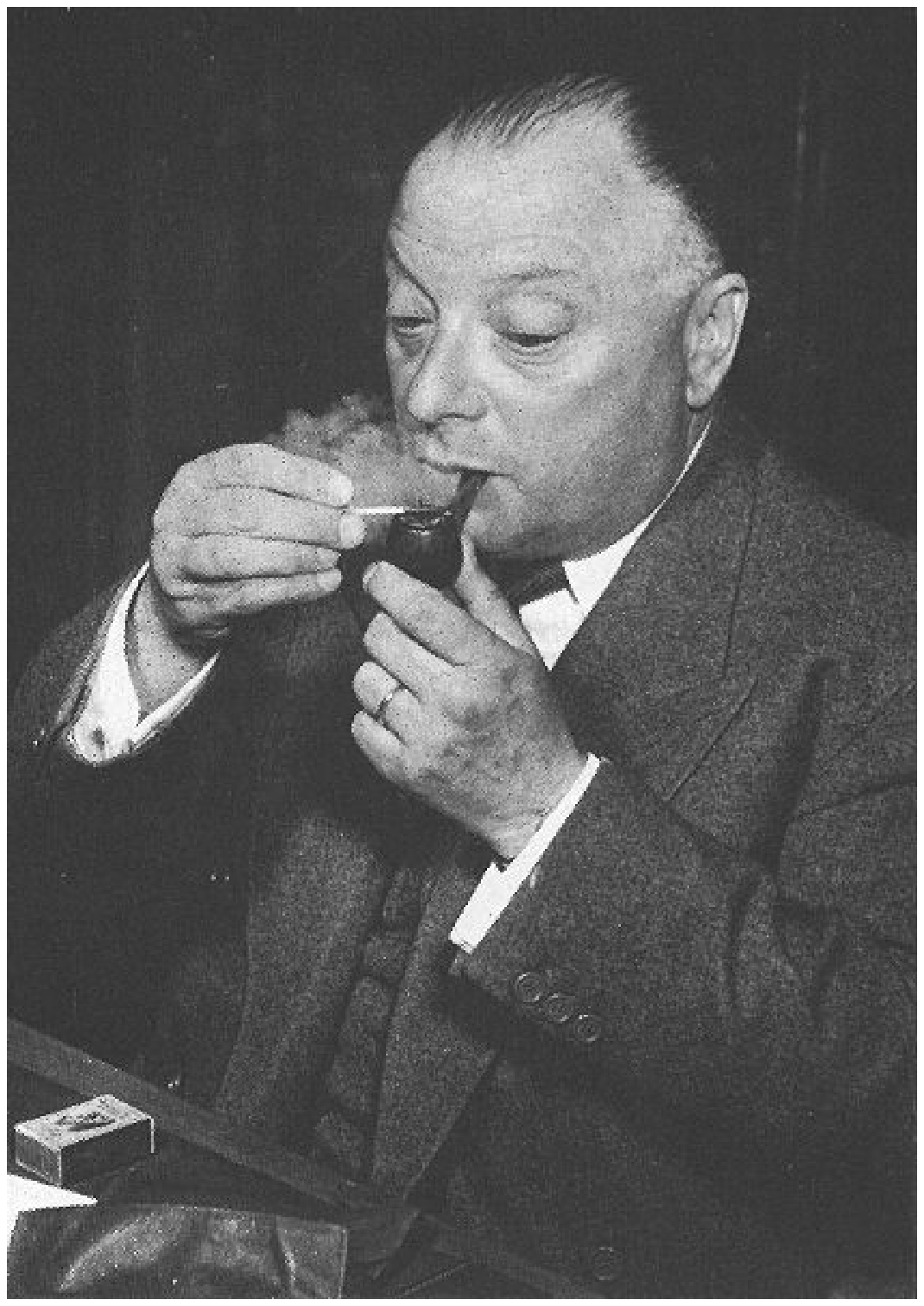}
}
\parbox{4cm}{
\epsfxsize=55mm
\hspace*{5mm}
\epsfbox{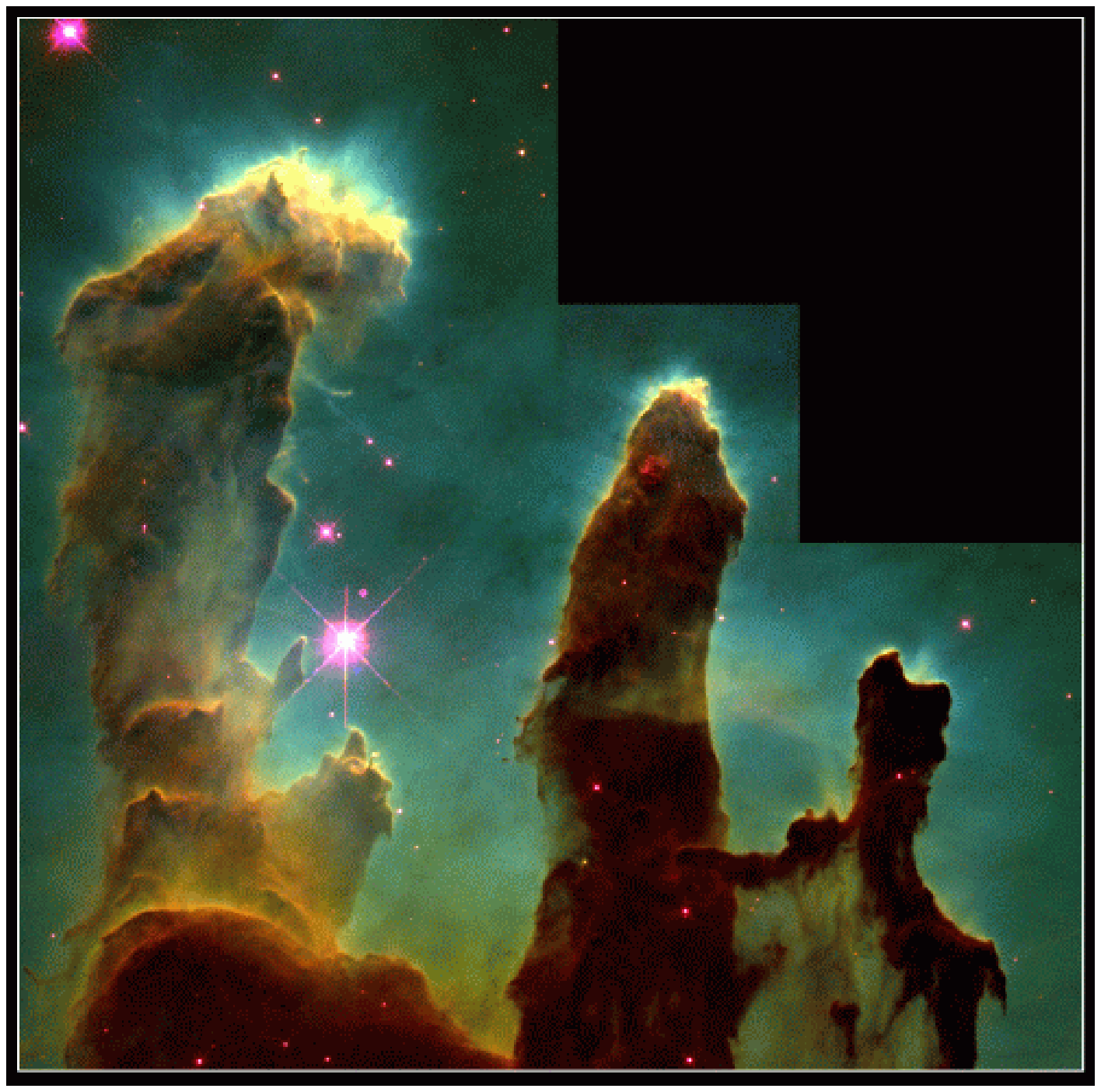}
}
\caption{\it The man who proposed it and an impression of its 
potential cosmological 
consequences: Wolfgang Pauli (nobel prize 1945, left panel   
\protect{\cite{bpauli}})
first thought 
about a neutral light particle being emitted in nuclear beta decays. 
Clouds of interstellar gas (right panel \protect{\cite{bstruct}}) 
act as a birthplace 
for new stars. Neutrinos may be important for the formation of structure 
in the early universe.
\label{wp}}
\end{figure}

\begin{itemize}

\item 
The neutrino seems to possess an at least million times smaller mass than the
lightest of the remaining particles, the electron.
While in the standard model the neutrino was introduced as massless 
``by hand'', this feature is especially problematic in unified theories, 
where 
the common treatment of neutrinos and charged fermions in extended multiplets
implies them to have (Dirac) mass terms of the same order of magnitude as 
the other fermions.

\item
Among all fundamental fermions the neutrino is the only one being
electrically uncharged. Thus the neutrino interacts a billion times
less often than an electron and may penetrate the entire earth without
even be deviated.
This is the reason why neutrinos, in spite of their 
tiny masses, may be that abundant that they contribute substantially to the 
mass of the universe, about twenty times more than the mass of all 
visible stars in the sky, and may influence the evolution of the universe,
e.g. the growth of structures, in a significant way.

\end{itemize}

\begin{figure}
\epsfxsize=100mm
\hspace*{1cm}
\epsfbox{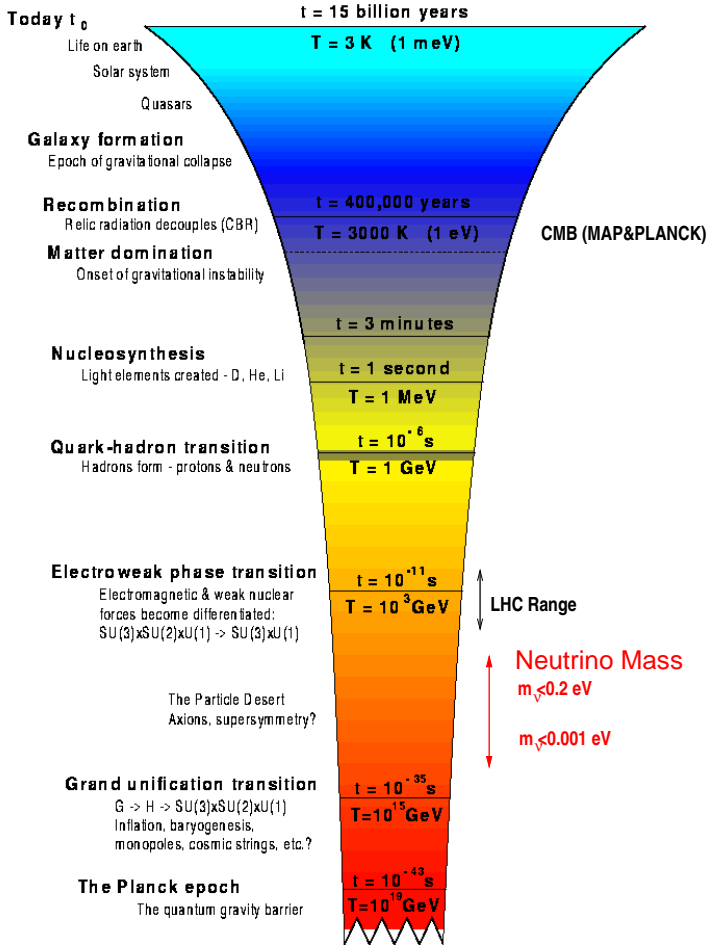}
\caption{\it The light neutrino mass can be related to a corresponding
large mass scale in unified theories, e.g. in the see-saw mechanism. 
This way the hunt for neutrino masses
offers a unique possibility to access physics at energy scales beyond the 
reach of running and future accelerators, which has been realized only 
shortly after the big bang. A neutrino mass of 0.1 eV - 0.001 eV corresponds,
via the mass of its heavy right-handed partner,
to energies of $10^{10}$ GeV or more, tiny fractions of a second after the big 
bang (background from \protect{\cite{bearl}}).
\label{right}}
\end{figure}

The basis for an understanding of these features relates them to each other
and was proposed in 1933 by the
Italian theoretician Ettore Majorana \cite{major}, 
one year before he dissappeared under
mysterious circumstances. Majorana found out that neutrinos, due to their
neutral charge, can be identical with their antiparticles, triggered by a new, 
so-called Majorana mass term
\footnote{In fact also pure usual ``Dirac'' mass terms for the neutrino
are possible but are disfavored in most fundamental theories.}. 
In 1979 T. Yanagida and independently
Murray Gell-Mann (nobel prize 1969), 
P. Ramond and R. Slansky found out that these 
additional Majorana mass
terms may cancel almost totally the usual Dirac mass terms  in the 
so-called ``see-saw mechanism''
\cite{seesaw}, yielding a natural explanation of the 
tiny neutrino masses. This would require the existence of  
right-handed heavy neutrinos as they are naturally predicted in 
``left-right-symmetric'' unified models. 
\footnote{Alternative mechanisms motivate neutrino masses at the weak scale,
a famous example is R-parity violating supersymmetry, see e.g. \cite{rpsusy},
where neutrino masses provide a window into deep relations of particles and 
forces. Also gravity induced non-renormalizable mass terms can play a role
in string-motivated scenarios, see e.g. \cite{cvetic}.
} 
The exact value of the mass then is correlated with
a higher energy scale predicted by the
underlying unified gauge group, and offers one of the rare possibilities to 
test these theories, since most of the predictions are observable only at very 
high energies, which are lying beyond the reach of present and future 
accelerators (see fig. \ref{right}).
The question to experimentalists thus remains: What is the mass of the
neutrino? The following review will outline the way to answer this question,
concentrating on two experimental approaches, yielding the complementary pieces
to solve the puzzle: Only {\it both} neutrino oscillations and neutrinoless
double beta decay {\it together} 
could solve this absolute neutrino mass problem.

\section{Neutrino oscillations}
\begin{figure}
\parbox{7cm}{
\epsfxsize=70mm
\epsfbox{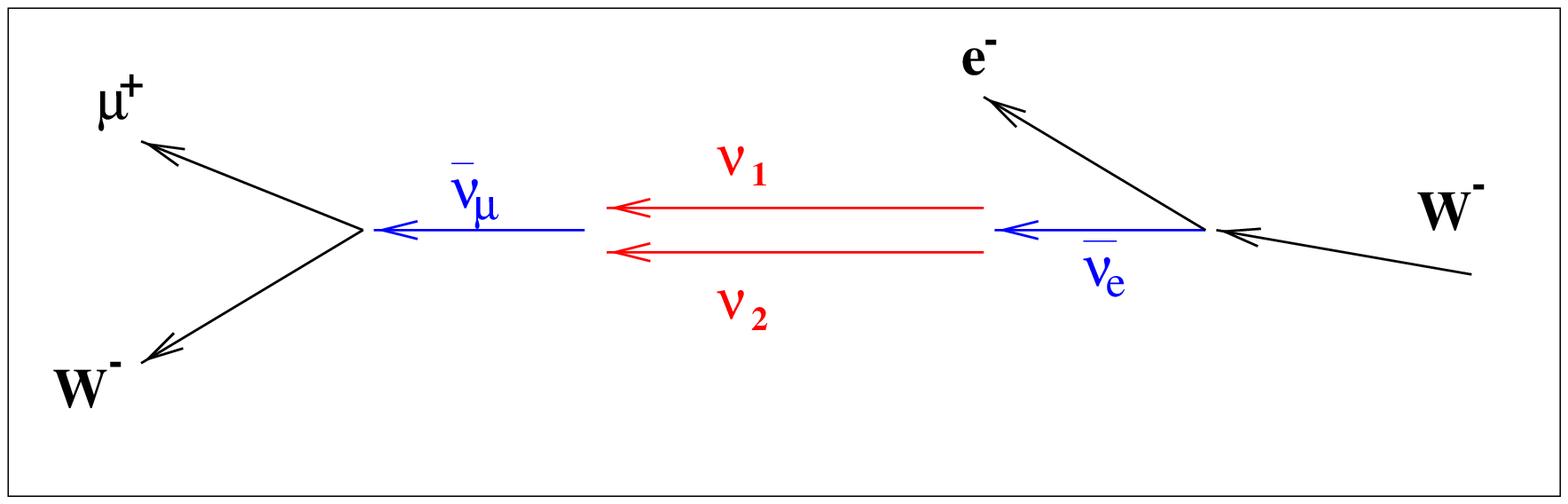}
}
\parbox{3cm}{
\epsfxsize=40mm
\epsfbox{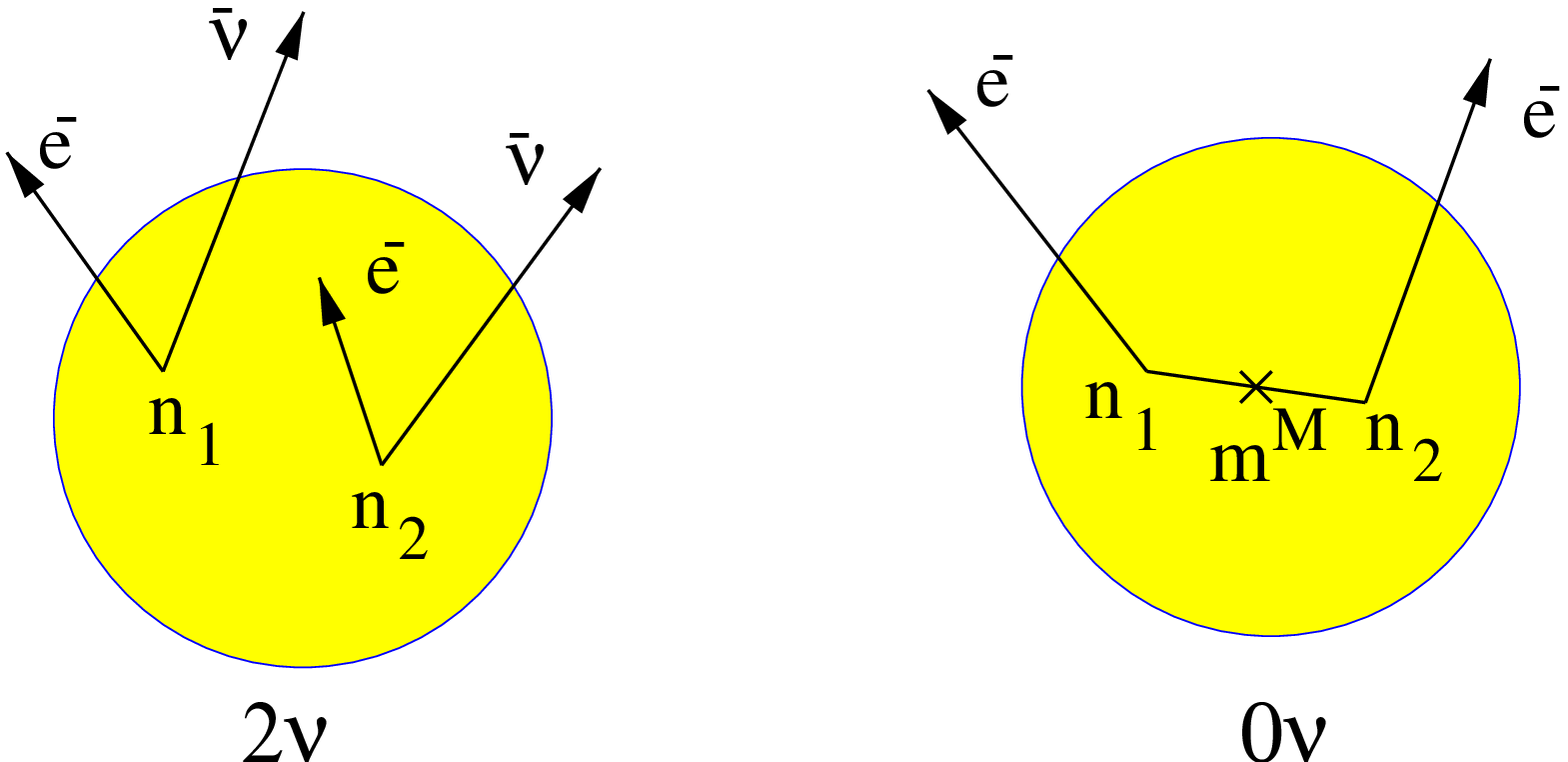}
}
\caption{\it Schematic representation of the two complementary processes
needed to solve the neutrino mass puzzle, neutrino oscillations and 
neutrinoless double beta decay. }
\label{schema}
\end{figure}

The fact that neutrinos are massive has finally been established by
neutrino oscillation experiments. 
Neutrino oscillations are a quantum mechanical process based on mixing
between the three neutrino flavors, which is possible if the flavor
(interaction) eigenstates $\nu_{\alpha}$ 
do not coincide with the mass eigenstates $\nu_i$. The flavor eigenstates
are thus given by a superposition of the mass eigenstates:
\be{flavor}
\nu_{\alpha} = \sum_{i=1}^3 U_{\alpha i} \nu_i
\label{flavor}
\ee
In that case
a neutrino, which is emitted as a flavor eigenstate $\nu_{\alpha}$
in a weak reaction, 
propagates as a superposition of the three mass eigenstates. If these
mass eigenstates are non-degenerate, they travel with 
different velocities and the composition in eq. (\ref{flavor}) is getting
out of phase. With a probability,
which is a function of the 
mass squared differences $\Delta m^2=m_i^2-m_j^2$ and the mixing 
$U_{\alpha i}$, after a certain distance the neutrino 
interacts as a different mass eigenstate $\nu_{\beta \neq 
\alpha}$ (see fig. \ref{schema}). 
Obviously neutrino oscillation experiments cannot give any information 
about the absolute mass scale in the neutrino sector, but yield informations
about mass (squared) differences, only.
Since the probability 
oscillates with
the propagation distance, this phenomenon, which was
predicted by Bruno Pontecorvo, 
after he disappeared in 1950 from England and later showed up again in
Russia, is called neutrino oscillations \cite{ponte}.

\begin{figure}
\epsfxsize=70mm
\hspace*{1cm}
\epsfbox{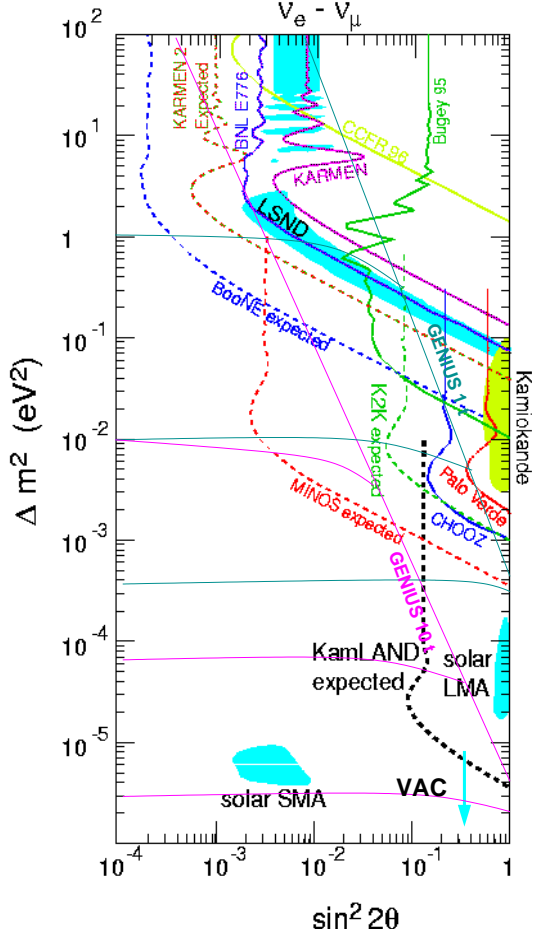}
\caption{\it 
Summary of neutrino anomalies. The solar neutrino deficit can be solved 
within the light blue regions of the small mixing and large mixing MSW 
oscillations, and the vacuum oscillations
(at $\Delta m^2\sim 10^{-9}-10^{-10}$ eV$^2$, not shown in the figure).
Only the large MSW solution is directly testable by KAMLAND. In the atmospheric
favored region (light green) $\nu_{\mu}-\nu_e$ oscillations are 
excluded by CHOOZ. 
The K2K and MINOS experiments will test for 
$\nu_{\mu} \rightarrow \nu_{\tau}$ oscillations. The  favored region of 
LSND is shown in light blue and can be tested in part by KARMEN and
in total by MINIBOONE. Also shown is the sensitivity of GENIUS 1 ton and 
GENIUS 10 t or comparable double beta decay projects
to $\nu_e \rightarrow \nu_{\mu,\tau,s}$ oscillations. The lines correspond 
to (from above) 
$m_{\nu_e}/m_{\nu_{\mu,\tau,s}}=\infty,0.01,0.1,0.5$
(background from \protect{\cite{blella}}). 
\label{fig4}}
\end{figure}

Up to now, hints for neutrino oscillations have been observed in solar and 
atmospheric neutrinos as well as the accelerator experiment LSND
(for an overview see fig. \ref{fig4}). 
\footnote{
It should be stressed that besides neutrino oscillations also new 
interactions beyond the standard model may provide solutions to  some of
the neutrino anomalies, see \cite{np}
}

\begin{itemize}

\item
A deficit of the number of solar neutrinos \cite{solan}
being expected has been 
confirmed in many experiments \cite{solar}
after the pioneering Chlor experiment \cite{chlor}
of Ray Davis in the Homestake mine. 
The oscillation mechanism of the solar $\nu_e$ in
(as normally assumed) $\nu_{\mu}$  \footnote{An alternative would be a 
fourth sterile $\nu_s$, see section 7}
may be induced via two different 
mechanisms. The usual neutrino oscillation mechanism  requires maximal 
mixing and suffers from the fact, that for this case the distance 
earth-sun has to be finetuned (vacuum oscillations). 
An alternative solution has been suggested by works of S. Mikheyev, 
Alexei Smirnov and L. Wolfenstein \cite{msw}:  
Resonant conversions, which are triggered by matter effects in the solar 
interior implying a level crossing of mass eigenstates, 
can cause the neutrino deficit. In this case both small as well as large 
mixing are allowed. The different solutions of the solar neutrino experiments 
correspond to different combinations of mass squared differences 
$\Delta m_{12}^2$ and mixing matrix elements $U_{12}^2$.
They will be tested by ongoing and future experiments such as
Super-Kamiokande, SNO and BOREXINO \cite{solnew} in the next years
\footnote{If one allows for larger confidence belts a third MSW ``LOW'' 
solution appears, which can be tested via its strong day-night effect at
low neutrino energies, observable at BOREXINO, LENS and the double beta 
and dark matter detector GENIUS (see below) \cite{sol2}.}.
Vacuum oscillations should lead to seasonal variations, the small mixing
MSW solutions should imply distortions of the energy spectrum and the
large mixing angle solution should show a small spectral distortion, 
a day-night effect of the total rate and a disappearance signal in the 
long baseline reactor experiment KAMLAND \cite{lbl} just under construction. 

\item
A similar effect has been observed in atmospheric neutrinos
\cite{atman}, which stem from 
the decay of the pions produced from cosmic ray interactions in the upper 
atmosphere and the following-up decays. Here Super-Kamiokande obtained a high 
precision result of a deficit of muon neutrinos compared to electron 
neutrinos. Even more convincing is the distortion observed for the zenith 
angle dependence of the muon neutrino flux, which provides a strong hint for 
$\nu_{\mu} \rightarrow \nu_{\tau}$
oscillations with maximal mixing and information about $\Delta m_{23}^2$
and $U_{23}^2$ \cite{superk}. 
Future long baseline experiments, K2K (already running), 
MINOS, and CERN-Gran Sasso \cite{lbl}, 
looking 
for oscillations in accelerator produced neutrino beams over distances of
several hundred kilometers will provide a check of this result 
by directly looking for $\nu_{\tau}$ appearance
and have the 
possibility to search for small contributions of 
$\nu_e \rightarrow \nu_{\tau}$ oscillations.

\item
Also an accelerator experiment, LSND, has 
reported evidence for $\nu_e-\nu_{\mu}$ neutrino oscillations. 
However, this evidence is generally understood as the most ambiguous.
The KARMEN experiment has excluded a large part of the favored region of LSND.
Since only two 
experimental evidences may be fitted with only three neutrinos.
the LSND result would require the existence of a fourth, sterile 
(i.e. not weakly interacting) neutrino (see section 7).
A decisive test will be obtained from the MINIBOONE experiment  
\cite{lsnd}. 

\end{itemize}

\section{Neutrinoless double beta decay}
Double beta decay ($0\nu\beta\beta$) corresponds to two single beta decays 
occurring in one nucleus and 
converts a nucleus (Z,A) into a nucleus (Z+2,A) (see fig. \ref{schema}).
While the standard model (SM) allowed process emitting two antineutrinos
\be{}
^{A}_{Z}X \rightarrow ^A_{Z+2}X + 2 e^- + 2 {\overline \nu_e}
\ee
is the rarest process observed in nature with half lives in the region of
$10^{21-24}$ years, more interesting is the search for 
the lepton number violating and thus SM forbidden neutrinoless mode,
\be{}        
^{A}_{Z}X \rightarrow ^A_{Z+2}X + 2 e^- 
\ee
which has been proposed by W.H. Furry in 1939 \cite{furry}.
In this case the neutrino is exchanged between the vertices
(see fig. \ref{schema}), a process being 
only allowed if the intermediate neutrino has a Majorana mass.
Neutrinoless double beta decay, when observed, also does not  
measure directly the neutrino mass. Since the neutrino in the propagator is only 
virtual, it does not have a definite mass. Propagating in the nucleus is 
the flavor eigenstate with the so-called effective neutrino Majorana mass
\be{}
\langle m \rangle = \left|\sum_j |U_{ej}|^2 e^{i\phi_j}m_j\right|,
\ee
which is a function of the mixing angles $U_{ej}$, complex phases $\phi_j$,
which allow for cancellations of the entering masses,  
and the neutrino mass eigenvalues. This quantity has exciting connections
to the observables in neutrino oscillation experiments. 
The most stringent limit on this quantity, $\langle m \rangle<0.35$ eV, 
is obtained by 
the Heidelberg--Moscow experiment \cite{hdmo}, 
which was initiated by one of the authors \cite{hdp} and is running since 
10 years in the Gran Sasso underground laboratory in Italy.
An impressive breakthrough to $10^{-2}-10^{-3}$ eV could be obtained 
realizing the GENIUS project proposed in 1997 \cite{genius}, 
a further proposal of H.V. Klapdor-Kleingrothaus,
operating 1-10 tons of enriched Germanium 
directly in a tank of 12 m diameter and height filled with liquid nitrogen.

\begin{figure}
\epsfxsize=80mm
\epsfbox{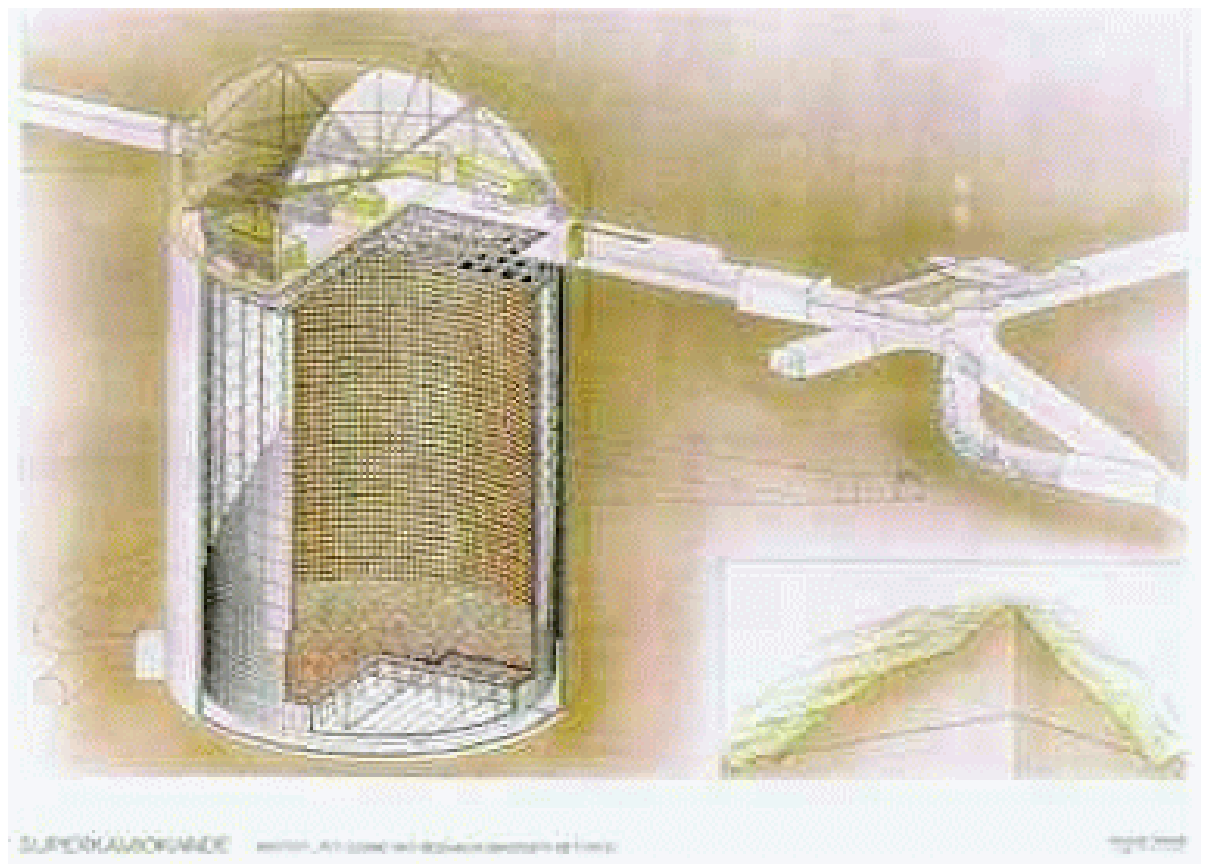}
\parbox{4cm}{
\epsfxsize=40mm
\epsfbox{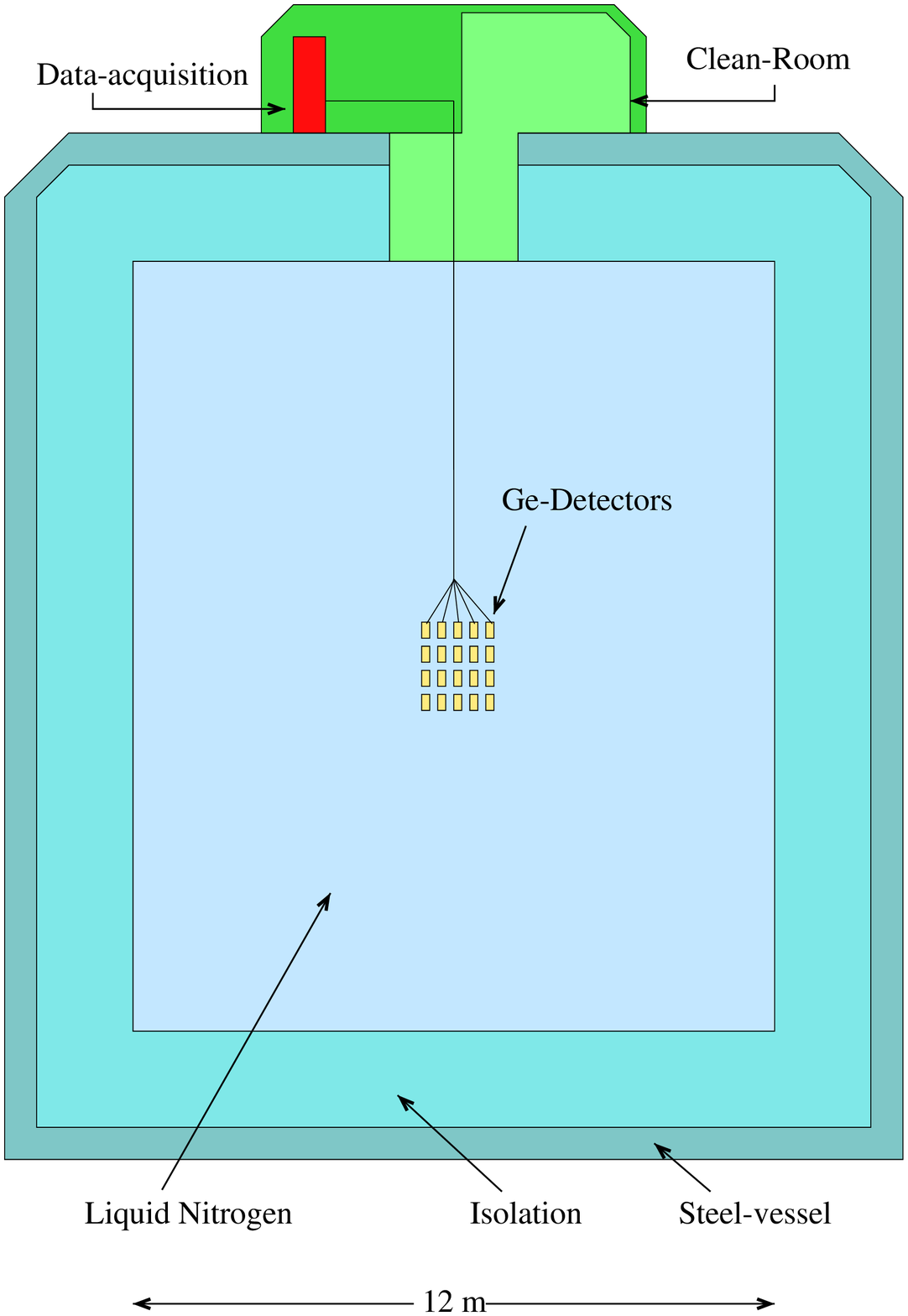}}
\parbox{4cm}{
\epsfxsize=40mm
\epsfbox{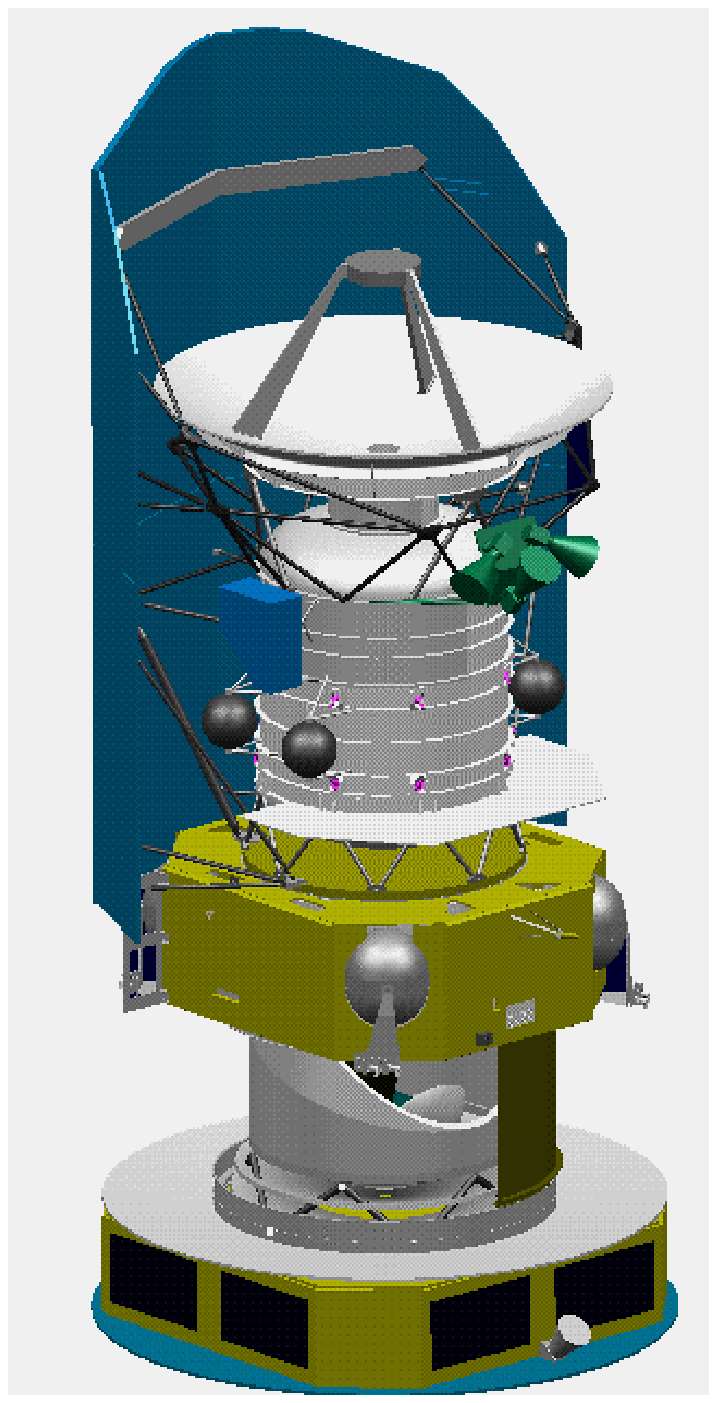}}
\caption{\it Schematic views of  
three experiments which will provide important complementary pieces of 
information about the neutrino mass. The Super-Kamiokande neutrino
oscillation experiment, the GENIUS double beta decay project 
(successor of the Heidelberg-Moscow experiment)
and the Planck cosmic microwave satellite \protect{\cite{bilder}}.
\label{exps}}
\end{figure}

How are the results in double beta decay and neutrino oscillations related?
In a recent work \cite{kps} the authors of this article in collaboration with 
Alexei Smirnov from the ICTP Trieste
were studying the relations of the neutrino oscillation 
parameters and the effective Majorana mass in the several possible 
neutrino mass scenarios 
and settled the conditions under which the neutrino mass 
spectrum can be reconstructed with future projects 
(see fig. \ref{exps}). 
In the following we will concentrate on
three extreme cases as examples, the hierarchical spectrum, 
the degenerate scheme and the inverse hierarchical scheme.

\section{Hierarchical schemes}

\begin{figure}
\epsfxsize=80mm
\epsfbox{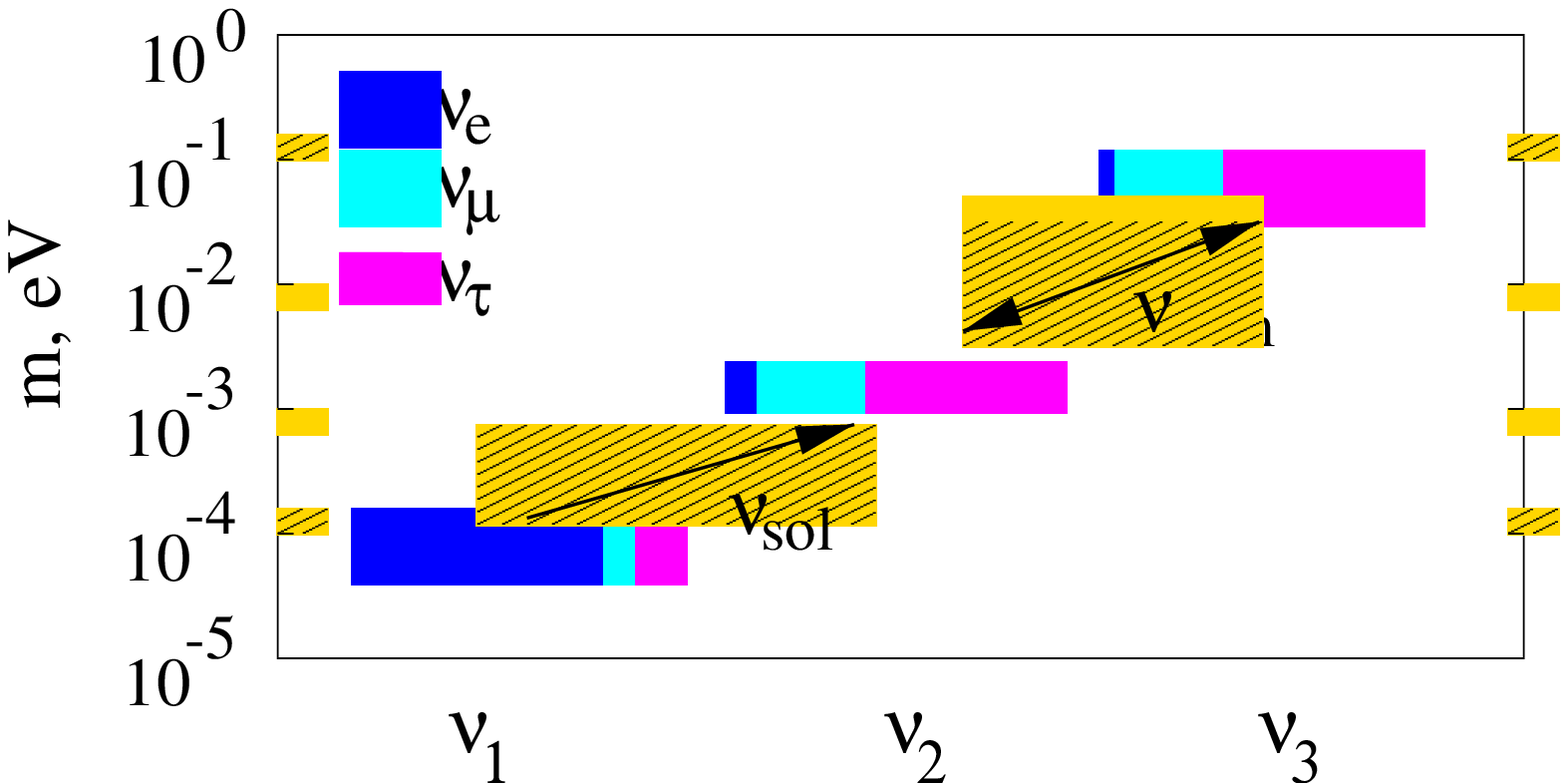}
\caption{Neutrino masses and mixings in the scheme with mass hierarchy,
shown is the example of small solar neutrino
mixing. Coloured bars correspond to flavor
admixtures in the mass eigenstates $\nu_1$, $\nu_2$, $\nu_3$. The quantity
$\langle m \rangle$ is determined by the dark blue bars denoting the admixture
of the electron neutrino $U_{ei}$.
\label{smi1}}
\end{figure}

Hierarchical spectra (fig. \ref{smi1})
\be{}
m_1 \ll m_2 \ll m_3
\ee
can be motivated by analogies with the quark sector and the simplest see-saw 
models. In these models the contribution of $m_1$ to the double beta decay 
observable $\langle m \rangle$
is small. The main 
contribution is obtained from $m_2$ or $m_3$, depending on the solution of the
solar neutrino deficit.

\begin{figure}[!ht]
\hspace*{1cm}
\epsfxsize=80mm
\epsfbox{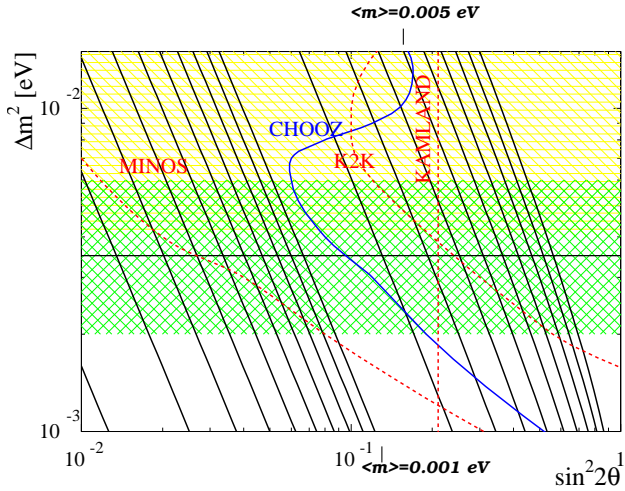}
\caption{\it 
Double beta decay observable $\langle m \rangle$ and oscillation parameters:
The case of hierarchical schemes with either the MSW small mixing solution 
or vacuum solution. Shown is the dominant
contribution of the third state to $\langle m \rangle$
which is constrained by the
CHOOZ experiment, excluding the region to the upper right. Further 
informations can be obtained from the long baseline project MINOS
and future double beta decay experiments \protect{\cite{kps}}.
\label{n1a}}
\end{figure}

If the small mixing angle solution is realized in solar neutrinos
(i.e. small $\nu_e-\nu_{\mu}$ mixing), 
the contribution of $m_2$ is small due to the small admixture $U_{e2}$.
The same is true for vacuum oscillations, where $U_{e2}$ is maximal but the 
mass of the second state is tiny. In these cases the main contribution 
to $\langle m \rangle$ comes 
from $m_3$. The contribution of the latter is shown in fig. \ref{n1a}. 
Here lines of constant $\langle m \rangle$
are shown as functions of the oscillation parameters $\Delta m_{13}^2$
and $U_{13}$, parametrized by $\sin^2 2 \theta_{13}$. 
The shaded areas show the mass $m_3 \simeq \sqrt{\Delta m^2_{13}}$
favored by atmospheric neutrinos with the horizontal line indicating 
the best fit value. The region to the upper right is excluded by the 
nuclear reactor 
experiment CHOOZ \cite{chooz}, 
implying $\langle m \rangle<2 \cdot 10^{-3}$ eV in the range favored by
atmospheric neutrinos. Obviously 
in this case only the 10 ton GENIUS experiment could observe a positive 
$0\nu\beta\beta$ decay signal. A coincidence of such a measurement with 
a signal of $\nu_e \rightarrow \nu_{\tau}$ oscillations at MINOS  
and a confirmation of solar vacuum or small 
mixing MSW oscillations by solar neutrino experiments would 
be a strong hint for this scheme.

\begin{figure}[!t]
\hspace*{1cm}
\epsfxsize=80mm
\epsfbox{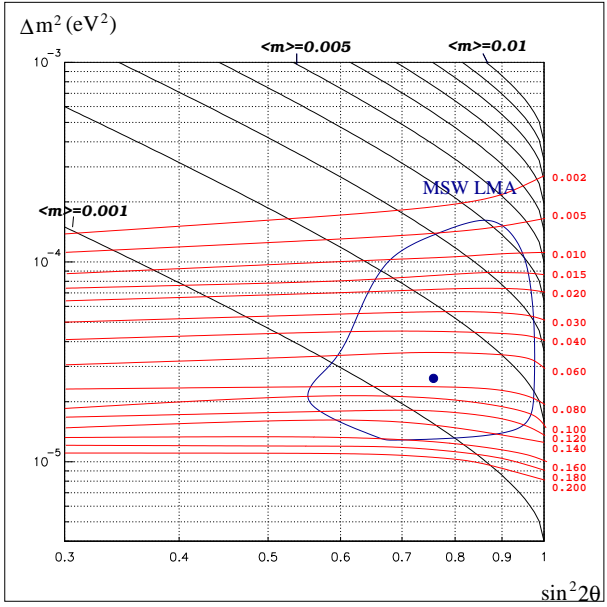}
\caption{\it 
Double beta decay oberservable $\langle m \rangle$ and oscillation parameters:
The case for the MSW large mixing solution of the solar 
neutrino deficit, where the dominant contribution to $\langle m \rangle$
comes from the second state. 
Shown are lines 
of constant $\langle m \rangle$ (diagonal) and constant day-night asymmetry 
(almost horizontal) \protect{\cite{kps}}.
The closed line shows the region allowed by present 
solar neutrino experiments. Complementary informations can be obtained
from double beta decay and the search for
a day-night effect in future solar neutrino experiments. 
\label{n2a} }
\end{figure}

If the large mixing solution of the solar neutrino deficit is realized,
the contribution of $m_2$ becomes large due to the almost maximal 
$U_{e2}$, now. Fig. \ref{n2a} shows values of $\langle m \rangle$ in the range of
the large mixing angle solution (closed line). The almost horizontal 
lines correspond to constant day-night asymmetries. A coincident measurement 
of $\langle m \rangle\simeq 10^{-3}$ eV, a day-night asymmetry of 0.07 at future 
oscillation experiments and a confirmation of the large mixing angle solution 
by KAMLAND would identify a single point in the large mixing angle MSW 
solution (in this example near the present best-fit point)
and provide a strong hint for this scheme.

\section{Degenerate schemes}

\begin{figure}
\epsfxsize=80mm
\epsfbox{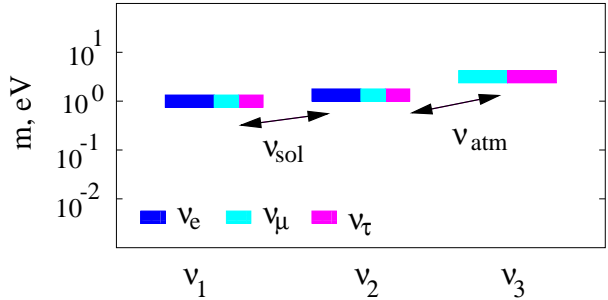}
\caption{Neutrino masses and mixings in the degenerate scheme, 
here the example with large solar neutrino mixing.
\label{smi2}}
\end{figure}

Degenerate schemes (fig. \ref{smi2})
\be{}
m_1 \simeq m_2 \simeq m_3 \simeq m_0
\ee
require a more general (and more complicated) 
form of the see-saw mechanism \cite{typeii}. 
One of their motivations is also, that  
a large overall mass scale allows neutrinos to be 
cosmologically
significant. Neutrinos with an overall mass scale of a few eV could play an 
important role as ``hot dark matter'' component of the universe.
When structures were formed 
in the early universe, overdense regions of (cold) dark matter provide the 
seeds of
the large scale structure, which later formed galaxies and clusters. A small
``hot'' (relativistic) component could prevent an overproduction of 
structure at small scales. 
Since structures redshift photons, this should imply also imprints 
on the cosmic microwave background (CMB), 
which could be measured by the future 
satellite experiments MAP and Planck \cite{cmb}. 

In degenerate schemes the mass differences are not significant. Since 
the contribution of $m_3$ is strongly bounded by CHOOZ again,
the main contributions to $\langle m \rangle$ come from $m_1$ and $m_{2}$.
The relative contributions of these states
depend on their admixture of the electron flavor, which is determined by the
solution of the solar neutrino deficit.

\begin{figure}[!t]
\hspace*{1cm}
\epsfxsize=80mm
\epsfbox{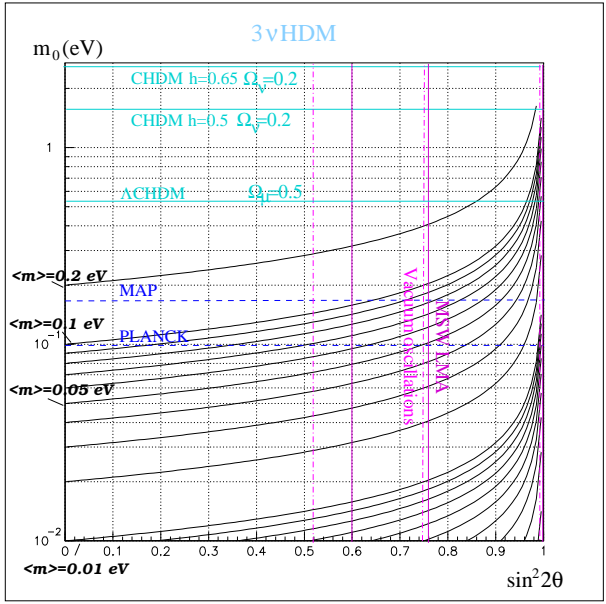}
\caption{\it 
Double beta decay oberservable $\langle m \rangle$ and oscillations parameters:
The case for degenerate neutrinos. Plotted on the axes are 
the overall scale of neutrino masses $m_0$
and the mixing $\sin^2 2 \theta_{12}$.
Allowed values for $\langle m \rangle$ for a given $m_0$ correspond to the 
regions 
between $m_0$ and the corresponding curved line. Also shown are informations 
which could be obtained from cosmological fits (see text) and the expected
sensitivity of the satellite experiments MAP and Planck. A value of 
$\langle m \rangle = 0.1$ eV in the case of small solar neutrino
mixing would be in 
the range to be explored by MAP and Planck \protect{\cite{kps}}.
\label{n3a}} 
\end{figure}

In fig. \ref{n3a} lines of constant double beta decay observables 
(solid curved lines) are shown together with information from 
cosmological observations about the overall mass scale (horizontal lines). 
Shown are best fits to the CMB and the large scale structure of Galaxy surveys
in different cosmological models as well as the sensitivity of MAP and Planck.
E.g., a $\Lambda$CHDM model with a total $\Omega_m=0.5$ of
both cold and hot dark matter as well 
as a cosmological constant, and a Hubble constant of $h=0.6$ would imply an 
overall mass scale of about 0.5 eV. However, the contributions of different 
mass eigenstates are in the same order of magnitude and may cancel, now.
Assuming a mixing corresponding to the 
best fit of solar  large mixing MSW or vacuum oscillations this yields 
$\langle m \rangle=0.2-0.5$ eV, just 
in the range of the recent half life limit of the 
Heidelberg--Moscow experiment. If even
larger mixing turns out to be realized in the solution of the solar 
neutrino deficit, this allows for a larger cancellation. 
A coincidence of the absolute mass scale reconstructed from double beta decay 
and neutrino oscillations with a direct measurement of the neutrino mass
in tritium beta decay spectra or its derivation from cosmological parameters 
determined from the CMB in the satellite experiments MAP and Planck would 
prove this scheme to be realized in nature. To establish this triple evidence 
however is difficult due to the 
restricted sensitivity of the latter approaches. Future tritium experiments
aim at a sensitivity down to $\cal{O}$(0.1~eV) and MAP and Planck have been
estimated to be sensitive to $\sum m_{\nu}=0.5-0.25$ eV.
Thus for neutrino mass scales below $m_0 < 0.1$ eV only a range for the 
absolute mass scale can be fixed by solar neutrino experiments and double beta 
decay.

\section{Inverse Hierarchy}

\begin{figure}
\epsfxsize=80mm
\epsfbox{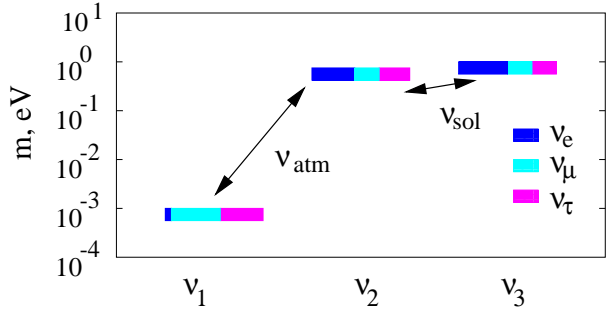}
\caption{Neutrino masses and mixings in the scheme with inverse hierarchy. 
Shown is the example with large solar neutrino mixing.
\label{smi3}}
\end{figure}

A further possibility is an inverse hierarchical spectrum (fig. \ref{smi3})
\be{}
m_3 \simeq m_2 \gg m_1
\ee
where the heaviest state with mass $m_3$
is mainly the electron neutrino, now. 

Its mass is mainly
determined by the atmospheric neutrinos, $m_3 \simeq \Delta m_{23}$.
Thus for the case of the small mixing angle solution one gets a unique 
prediction of $\langle m \rangle=(5-8) \cdot 10^{-2}$ eV, which could be tested by the 
1 ton version of GENIUS. For the vacuum or large mixing
MSW solution cancellations of the two heavy states become possible
and $\langle m \rangle<8 \cdot 10^{-2}$ eV. A test of the inverse hierarchy 
is possible 
in matter effects of neutrino oscillations. For this case the MSW
level crossing 
happens for antiparticles rather than for particles. Effects could be 
observable in long baseline experiments and in the neutrino spectra of 
supernovae \cite{supn}.

\section{Four neutrinos}
Sterile neutrinos, which do not couple to the weak interactions, 
can easily be motivated in superstring inspired models:
multidimensional candidates for a final ``Theory of Everything'', in which the 
fundamental constituents of matter have a string rather than a particle 
character. Such theories could accomodate for additional neutrinos in
different ways.
Examples are extended gauge groups, fermions living in extra 
(compactified) dimensions
as well as a mirror world,
which contains a complete duplicate of matter and forces building the universe,
interacting only via gravity.
In the latter case $\langle m \rangle=0.002$ eV is predicted \cite{sterile}.  

\begin{figure}
\epsfxsize=80mm
\epsfbox{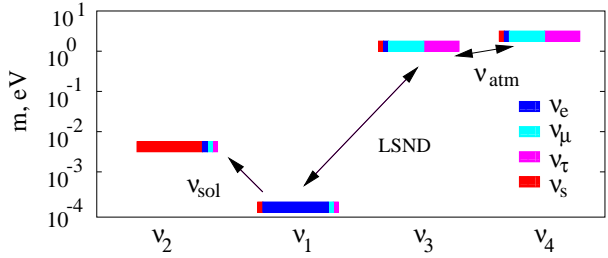}
\caption{Neutrino masses and mixings in the four neutrino scheme, shown is the 
example with small solar neutrino mixing.
\label{smi4}}
\end{figure}

If the four neutrinos are arranged as two pairs of degenerate states 
(mainly $\nu_e-\nu_s$ for solar and $\nu_{\mu}-\nu_{\tau}$ 
for atmospheric neutrinos)
separated by a LSND gap, all three neutrino anomalies can be solved and the 
two heavy states can account for the hot dark matter. The main contribution
to $\langle m \rangle$ comes from the heavy states, then, and can be derived from the 
LSND result. Depending on the phase of these two contributions $\langle m \rangle$ can be
as large as ${\cal O}(10^{-3}$ eV). A strong hint for the scheme would be a 
coincidence of the $\Delta m^2$ favored in LSND and possibly 
MINIBOONE, cosmological 
observations and double beta decay, together with the discovery of sterile 
neutrinos in solar neutrino oscillations by SNO.

\section{Summary}

The recent years brought exciting developments in neutrino physics.
Neutrino oscillations have finally been confirmed in atmospheric neutrinos 
and at the same time double beta decay experiments realized for the first time
a sensitivity, leading to strong implications on the neutrino mass spectrum
and cosmological parameters.
After this
particle physics now seems to enter its ``neutrino epoche'': The neutrino
mass spectrum and its absolute mass scale offer unique possibilities 
to provide crucial information
for cosmology and theories beyond the standard model. Only both neutrino
oscillations and neutrinoless double beta decay together
have the chance to solve this 
neutrino mass problem (see also, e.g. \cite{znbb})
and to set the absolute scale in the neutrino sector:
If the solution of the solar neutrino deficit and the character of hierarchy
(direct or inverse)
is determined in neutrino oscillation experiments, the following informations 
will be obtained from a future double beta decay project:

For the case of direct/normal hierarchy, a confirmation of the 
small mixing MSW solution would mean: If double beta decay would 
be measured with $\langle m \rangle>0.1$ eV this would establish a degenerate spectrum
with a fixed mass scale. If $\langle m \rangle$ is measured in the range 
$(0.5-3)\cdot 10^{-2}$ eV a partially degenerate spectrum,
$m_1 \simeq m_2 \ll m_3$, with fixed mass scale
is realized in nature. For $\langle m \rangle<2 \cdot 10^{-3}$ eV a hierarchical 
spectrum exists in nature. For the large mixing MSW solution a value of
$\langle m \rangle>3 \cdot 10^{-2}$ eV implies a degenerate spectrum with a region 
for the mass scale determined by the solar mixing angle. For 
$\langle m \rangle< 2 \cdot 10^{-2}$ eV a partially degenerate or hierarchical 
spectrum is realized in nature and a region for the mass scale is set 
by the solar mixing angle. If $\langle m \rangle<2 \cdot 10^{-3}$ eV is measured
the spectrum is hierarchical. If vacuum oscillations are the correct
solution for the solar neutrino deficit a value of 
$\langle m \rangle> 3 \cdot 10^{-2}$ eV implies degeneracy, $\langle m \rangle> 2 \cdot 10^{-3}$ eV
partial degeneracy and $\langle m \rangle< 2 \cdot 10^{-3}$ eV hierarchy, but no 
information about the absolute mass scale is obtained.

For the case of inverse hierarchy the situation is more predictive. 
For the small mixing angle MSW solution $\langle m \rangle \equiv (5-8)\cdot 10^{-2}$ eV is expected.
For large mixing angle MSW or vacuum oscillations 
one awaits
$\langle m \rangle < 8 \cdot 10^{-2}$,
above this value the scheme approaches the degenerate case.

In four neutrino schemes $\langle m \rangle$ can be as large as
${\cal O}(10^{-3})$ eV. A conincidence of a double beta decay signal with
the $\Delta m^2$ favored in LSND and possibly in MINIBOONE, an imprint of 
neutrinos as hot dark matter in the CMB as well as the discovery of sterile 
neutrinos in SNO would prove the scheme and fix the mass scale.

This outcome will be a large step both towards the understanding 
of the evolution of the universe and towards the dream of a 
unified theoretical 
description of nature.
We are entering an exciting decade!

\end{document}